\documentclass{article}

\usepackage{microtype}
\usepackage{graphicx}
\usepackage{booktabs} %
\usepackage{listings}
\usepackage[scaled=0.8]{beramono} %
\usepackage{amsmath,amssymb,amsfonts}

\usepackage{hyperref}

\usepackage{cleveref}
\usepackage{subcaption}
\usepackage{xspace}
\usepackage{multirow}
\usepackage[table]{xcolor}

\Crefname{line}{L\hspace{-2pt}}{L\hspace{-2pt}}

\newcommand{\sysname}{JaxPP\xspace}

\usepackage[accepted]{mlsys2025}

\mlsystitlerunning{Scaling Deep Learning Training with MPMD Pipeline Parallelism}

\begin{document}

\twocolumn[
\mlsystitle{Scaling Deep Learning Training with MPMD Pipeline Parallelism}

\mlsyssetsymbol{equal}{*}

\begin{mlsysauthorlist}
\mlsysauthor{Anxhelo Xhebraj}{nvda}
\mlsysauthor{Sean Lee}{nvda}
\mlsysauthor{Hanfeng Chen}{nvda}
\mlsysauthor{Vinod Grover}{nvda}
\end{mlsysauthorlist}

\mlsysaffiliation{nvda}{NVIDIA, USA}

\mlsyscorrespondingauthor{Anxhelo Xhebraj}{axhebraj@nvidia.com}
\mlsyscorrespondingauthor{Sean Lee}{selee@nvidia.com}
\mlsyscorrespondingauthor{Hanfeng Chen}{hanfengc@nvidia.com}
\mlsyscorrespondingauthor{Vinod Grover}{vgrover@nvidia.com}

\mlsyskeywords{Large-Language Models, Distributed Machine Learning, Pipeline Parallelism,
Single-Program Multiple-Data, Multiple-Program Multiple-Data}

\vskip 0.3in

\begin{abstract}
We present \sysname, a system for efficiently scaling the training of large deep learning
models with flexible pipeline parallelism.
We introduce a seamless programming model that allows implementing user-defined pipeline
schedules for gradient accumulation.
\sysname automatically distributes tasks, corresponding to pipeline stages, over
a cluster of nodes and automatically infers the communication among them.
We implement a MPMD runtime for asynchronous execution of SPMD tasks.
The pipeline parallelism implementation of \sysname improves hardware utilization by up
to $1.11\times$ with respect to the best performing SPMD configuration.

\end{abstract}
]

\printAffiliationsAndNotice{}  %

\section{Introduction}\label{sec:introduction}
The capability of deep learning models in a wide array of tasks has been shown to scale with model
size~\cite{brown_language_2020-1, dosovitskiy_image_2021}.
Consequently, researchers are training increasingly larger models~\cite{shoeybi_megatron-lm_2020,chowdhery_palm_2022}.
Considerable development efforts are required to run such experiments, which are often justified only for
well-performing models, thus restricting the exploration of models that excel when scaled~\cite{hooker_hardware_2020}.

A primary challenge in developing large models lies in their efficient parallelization across
various hierarchies (cores, devices, hosts, data centers) to maximize resource utilization and
minimize device communications
and tensor layout changes~\cite{shoeybi_megatron-lm_2020, narayanan_efficient_2021, fedus_switch_2022,
pope_efficiently_2022}.
Early works concentrated on manually re-implementing models to run them across multiple devices, which resulted in
highly optimized runtimes for specific models~\cite{shoeybi_megatron-lm_2020, narayanan_efficient_2021}.
However, this approach is time-consuming and error-prone.

The GSPMD programming model, as implemented in the XLA compiler~\cite{xu_gspmd_2021}, simplifies the parallelization
of linear algebra workloads.
It requires only lightweight annotations that specify how tensors in a computation should be sharded across a mesh of
devices, with the compiler automatically handling the placement of collective operations for communication.
Once the tensors are sharded and collective operations are in place, the computation is carried out in a
Single-Program Multiple-Data~(SPMD) fashion.
This decoupling of sharding annotations from computation definitions facilitates experimentation with various
intra-operator parallelism strategies~\cite{zheng_alpa_2022} such as data parallelism and tensor parallelism to minimize latency~\cite{fedus_switch_2022, pope_efficiently_2022}.

Despite GSPMD's near-ideal solution, the SPMD model works well in practice only when high-bandwidth
links connect accelerators, e.g. NVSwitch for GPUs and ICI for TPUs.
This is because the collective operations necessary for the SPMD computations
stress the network's bandwidth.
It is well-known that scaling high-bandwidth links to larger device meshes quickly becomes infeasible.
For example scaling the training of LLMs on TPUs~\cite{chowdhery_palm_2022} required designing a separate
system~\cite{barham_pathways_2022} to extend the SPMD model to cross low-bandwidth (DCN) domains.
In settings where device connection has a low bandwidth, communication overhead can be greatly reduced with
pipeline parallelism~\cite{huang_gpipe_2019}, which requires only Point-to-Point (P2P) communication.

GSPMD can implement only one variant of pipeline parallelism, precluding any form of pipeline
parallelism that requires a Multiple-Program Multiple-Data (MPMD) paradigm.
This limitation restricts significantly the types of computations that can be pipelined, and
precludes various pipeline schedules that improve throughput and memory usage.
In practice, best performing training configurations~\cite{shoeybi_megatron-lm_2020} use a mix of pipeline, tensor, and
data parallelism.
Tensor parallelism is mapped over the high-bandwidth mesh dimension while data and pipeline parallelism are mapped
over the low-bandwidth dimension.

Our work introduces \sysname, a system for distributed training of large models.
Unlike other systems such as Megatron~\cite{shoeybi_megatron-lm_2020} and DeepSpeed~\cite{smith_using_2022},
model implementations do not have to commit to a concrete parallelization strategy.
Instead, by building on top of GSPMD, parallelism is decoupled from the implementation and is introduced through
lightweight sharding annotations.
Additionally, \sysname advances beyond SPMD by allowing arbitrary MPMD distributed dataflow in the
pipeline parallelism dimension.
We make the following contributions:
\vspace{-6pt}
\begin{itemize}
  \setlength\itemsep{2pt}
  \item We introduce a novel programming model that enables users to express pipeline parallelism seamlessly.
    The programming model does not require any user intervention to handle (potentially non-adjacent)
    communication across pipeline stages.
  \item We present a task-graph implementation that enables \sysname to schedule tasks
    over a distributed mesh of devices, infer communication among them, and perform
    resource management tasks such as allocation and buffer deletion.
  \item We present \sysname's single-controller MPMD runtime, that supports
    the execution of arbitrary user specified pipeline schedules.
  \item We demonstrate the benefits of our design by highlighting performance characteristics of
    \sysname and compare it against state of the art alternatives on practical large-scale training
    benchmarks.
\end{itemize}
\vspace{-6pt}
The rest of the paper is structured as follows.
In \Cref{sec:motivation}, we motivate our system by describing the limitations of
existing parallelization ``interfaces'' when applied to pipeline parallelism.
Then, we give an overview of \sysname~(\Cref{sec:design}) and describe its runtime~(\Cref{sec:sched}).
We extensively evaluate \sysname's performance characteristics and compare against
other state of the art systems~(\Cref{sec:evaluation}).
We conclude by highlighting related work~(\Cref{sec:related}) and final
remarks~(\Cref{sec:conclusion}).

\section{Motivation: User-Driven Partitioning Beyond SPMD}\label{sec:motivation}

Scaling deep learning model training to large distributed clusters of devices requires a combination
of several parallelization techniques.
For example, the training of Llama 3 models~\cite{dubey2024llama3herdmodels}
used the combination of data parallelism, tensor parallelism, pipeline parallelism, and context parallelism.
Implementing these parallelization strategies manually requires substantial effort and expertise.

To simplify this process, a few libraries and frameworks enable the post-hoc parallelization of
existing models without major rewrites.
Notably, recent work such as GSPMD~\cite{xu_gspmd_2021}, which is integrated within
JAX~\cite{jax2018github}, and PartIR~\cite{alabed_partir_2024}
introduce programming models that decouple model implementation from parallelization strategies,
making distributed training more accessible.
We now briefly describe JAX's programming model for parallel computations and explore its
limitations for pipeline parallelism.

\begin{figure*}
  \centering
  \begin{minipage}[b]{.43\textwidth}\footnotesize
    \begin{lstlisting}[numbers=none]
$\text{@}$shard(("batch", "emb"), ("emb", "mlp"), ("mlp", "emb"))
def ffn($X$, $W^{(1)}$, $W^{(2)}$):
  $H^{(1)}$ = relu($X W^{(1)}$)
  $H^{(1)}$ = shard($H^{(1)}$, ("batch", "mlp"))
  $H^{(2)}$ = $H^{(1)} W^{(2)}$
  return shard($H^{(2)}$, ("batch", "emb"))
    \end{lstlisting}
    \subcaption{Model implementation with named axes}\label{fig:ffn}
    \begin{align*}
      &\textsf{partitioning} = [ \\
      &  \quad \textsf{\textbf{batch}} \triangleright \textsf{data}, \\
      &  \quad \textsf{\textbf{mlp}} \triangleright \textsf{model} \\
      &]
    \end{align*}
    \subcaption{Partitioning specification}\label{fig:partitioning}
  \end{minipage}
  \begin{minipage}[b]{.55\textwidth}\footnotesize
    \begin{center}
      \includegraphics[width=.8\textwidth]{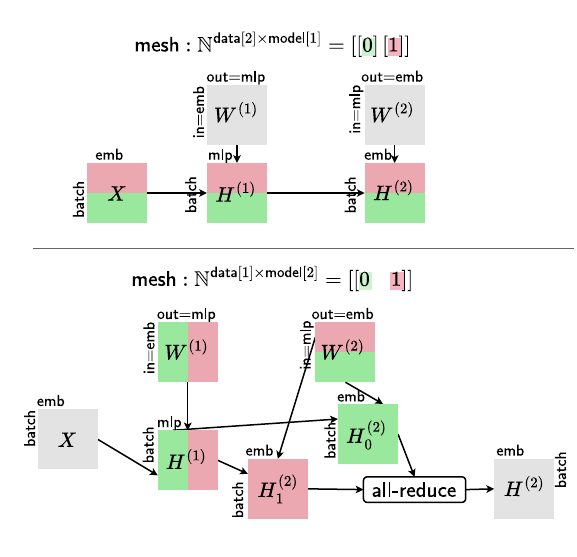}
      \subcaption{Different parallelism instantiations depending on the mesh shape}\label{fig:parallel}
    \end{center}
  \end{minipage}
  \caption{
    \textbf{Configurable Parallelism Through Named Axes in JAX}~\cite{jax2018github}
    Top left~(\ref{fig:ffn}): Model implementation where array axes are annotated with logical names.
    Bottom left~(\ref{fig:partitioning}): Partitioning specification mapping logical axis names to mesh axes.
    Right~(\ref{fig:parallel}): Two parallel instantiation, data-parallel on the top with mesh shape
    \lstinline|[("data", 2) ("model", 1)]|
    while a tensor-parallel implementation at the bottom when the mesh shape is \lstinline|[("data", 1) ("model", 2)]|.
  }\label{fig:ffn-full}
\end{figure*}

\subsection{SPMD Parallelization Through Named Axes}\label{subsec:named_axes}
To parallelize an array computation in JAX, we arrange a set of devices in a
logical mesh, which is a multi-dimensional array of non-repeating devices.
The mesh shape and device order can be arbitrary, but it is usually such
that dimensions of the mesh correspond to a particular communication
bandwidth.
For example, 4 nodes with 8 GPUs each could have a mesh shape of (4, 8)
where each row corresponds to the 8 devices present on each node.
Therefore communication between devices within the same row is
faster over communication between devices across different rows.
The mesh dimensions can also be named, for example, \lstinline|[("data", 4) ("model", 8)]|.

Given a mesh, an array can be \emph{sharded} (or partitioned) by mapping
some axes of the array to some axes of the mesh.
If an array axis is not mapped to any mesh axis, then that axis
is replicated across the remaining dimensions of the mesh.
The snippet below shows shardings of the two-dimensional array \lstinline|A| arising due
to different partitioning specifications.

\begin{lstlisting}
  # mesh.shape=[("data", 4) ("model", 8)]
  # A.shape=(n, m)
  shard(A, (None   , "model")) # col (n  , m/8)
  shard(A, ("data" , None))    # row (n/4, m  )
  shard(A, ("data" , "model")) # 2D  (n/4, m/8)
\end{lstlisting}

The examples above mentioning only one axis of the mesh, will lead to the replication
of the sharded tensor on the unmentioned axis of the mesh.
For example, in the first case that mentions only the \lstinline|"model"| axis,
\lstinline|A| is replicated across the 4 \lstinline|"data"| groups and sharded
column-wise within each of them.

Instead of specifying concrete mesh axes in model definitions,
usually \emph{logical axis names} (\emph{named axes}) are used.
This allows exploring several parallelization strategies
with different mesh shapes without any changes to the model implementation.
Axis names must be unique for the axis of one array, but can be shared across multiple arrays.

\Cref{fig:ffn} shows the definition of a Feed-Forward Network~(FFN) using logical axis names.
Note that the function has no collectives in its implementation and can be run on a single device.
The function takes as argument a 2D input $X$ with \lstinline{batch} and \lstinline{emb} (embedding)
dimension and maps it to the 2D output $H^{(2)}$ with same logical axis names.
The body of the function consists of the application of two parameters $W^{(i)}$ interleaved
with an activation.
Note that, while the input and output share logical axis names, their sizes can differ.
\Cref{fig:partitioning} specifies on what mesh dimension each logical axis name should
be sharded on.
The parallelization of the computation is still undefined and will depend only on
the concrete instantiation of the device mesh.

\textbf{Data Parallelism (DP)} replicates the model weights across devices while
partitioning the batch among them.
\Cref{fig:partitioning}~(Top) shows the corresponding mesh instantiation of shape
\lstinline|[("data", 2) ("model", 1)]| to achieve this.
Since all axes of the weights $W^{(i)}$ are either unbound (\lstinline|emb|) or are mapped
to a mesh dimension of size 1 (\lstinline|mlp| mapped to \lstinline{model}),
the weights are replicated (shown as gray blocks) across the two devices.
For training, each ``replica'' computes the gradients with respect to the local
batch.
The gradient computation, which is not shown in the figure, requires contracting the
activations on the batch dimension, leading to an \lstinline|all_reduce| operation
(replicas synchronize their gradients).
This parallelization strategy allows training over larger ``global'' batch sizes, potentially
leading to faster convergence.

\textbf{Tensor Parallelism (TP)}~\cite{shoeybi_megatron-lm_2020} partitions weights of an individual
layer over multiple devices.
This allows running large models for which the training state does not fit into a single device.
However, depending on the operations performed on the weights, collective operations may be required
to complete the computation. XLA inserts them automatically as needed.
The ``Megatron-style'' parallelization strategy corresponding to the mesh shape
\lstinline|[("data", 1) ("model", 2)]| is shown in \Cref{fig:parallel} (Bottom).
The subscript of a variable (e.g., $0$ in $H_0$) denotes the \emph{shard}.
Because the \lstinline{mlp} axis is bound to the \lstinline|model| axis of the mesh, which is composed
of 2 devices, $W^{(1)}$ is partitioned on the output dimension (column-wise)
while $W^{(2)}$ is partitioned in the input dimension (row-wise).
The input and output of the function are replicated.
The collective necessary for performing the parallel computation is inserted implicitly
by XLA's SPMD partitioner.
The second matrix-multiply operation $H^{(1)} W^{(2)}$ requires only one final
\textsf{all-reduce} to compute the replicated output.

It is possible to combine DP and TP by defining a larger mesh such as \lstinline|[("data", 4) ("model", 8)]|.
In this scenario, 32 GPUs are split into 4 DP ``groups'' each constituted by 8 TP groups.
Weights are replicated across the 4 DP groups and sharded across the 8 TP groups within each DP group.
Similarly the batch is sharded across the 4 DP groups and then each shard is replicated within each TP
group.

Finally, this programming model also allows more complex parallelism strategies such as
Expert Parallelism (EP)~\cite{lepikhin_gshard_2020}, where expert weights and intermediate activations
are sharded and multiplied in parallel.

\subsection{Limitation of SPMD: Pipeline Parallelism}\label{subsec:pp}
All the parallelism strategies described so far fall into the SPMD category.
Under this model, a single program is compiled and executed across multiple devices, each processing distinct
input shards.
This approach enables scalable deployment across thousands of partitions and simplifies scheduling,
especially for collective operations.
However, for larger scale of number of devices, collectives necessary in the SPMD model can greatly harm performance.

Pipeline Parallelism (PP) offers an alternative by introducing temporal parallelism, dividing the computation
graph into \emph{stages}, and performing \emph{gradient accumulation} over smaller
partitions of the batch, called \emph{microbatches}.
From here, we use the term \emph{actor} to refer to a group of devices executing
the same logic.

\subsubsection{The Importance of Pipeline Schedules}\label{subsec:ppsched}
The first successful application of pipeline parallelism with synchronous gradient application was
demonstrated in GPipe~\cite{huang_gpipe_2019}.
A neural network's layers are split into \emph{stages}.
For each stage there is a forward computation and backward computation which must be scheduled
in the same actor as the forward one.
Each actor is assigned one stage, and iteratively executes the forward computation for each microbatch
by, first saving potential activations needed for the backward computation, and then sending the output
to the next actor.
At the end of all microbatches, an actor receives the gradients
of the activations from the succeeding actor and executes each backward computation.
Finally, the accumulated gradients are used to update the model weights and optimizer
parameters at the end of the training step.
Since the activations of each microbatch have to be stored until the corresponding backward computation,
memory usage in GPipe is proportional to the number of microbatches.
Therefore, GPipe is usually combined with activation rematerialization~\cite{DBLP:journals/corr/ChenXZG16}.

Later works such as 1F1B~\cite{narayanan_pipedream_2019} and Interleaved 1F1B~\cite{narayanan_efficient_2021}
improved both memory usage and throughput of PP.
\Cref{fig:gpipe_1f1b} shows the difference between the two schedules.
The key realization is that the various stages of different gradient accumulation iterations
can be scheduled arbitrarily as long as data dependencies are honored.
Therefore, a gradient accumulation loop can be implemented in various ways
with different \emph{schedules} which describe the order and on which actor
each stage's computation (or \emph{task}) is run.

The 1F1B schedule shortens the lifetime of activations by eagerly scheduling the execution of backward stages.
As a result, memory requirements become proportional to the number of stages instead of
the number of microbatches, potentially translating to a $2\times$--$3\times$ reduction in
activation memory.
This increased memory availability also improves throughput since more activations
can be stored and not rematerialized.
Consequently, the end-to-end time of a training step can be reduced by $20\%$
as we explain in \Cref{subsec:spmd_vs_jaxpp}.

Interleaved 1F1B further reduces idling time by assigning multiple stages to each actor.
The number of stages per actor is referred to as the degree of \emph{circular repeat}.
As the degree of circular repeat increases, stages become smaller, enabling finer-grained scheduling.
This approach improves throughput, but introduces additional communication overhead.

\subsubsection{SPMD Encoding of Pipeline Parallelism}\label{subsubsec:gspmdpp}

\begin{figure}%
  \includegraphics[width=\linewidth]{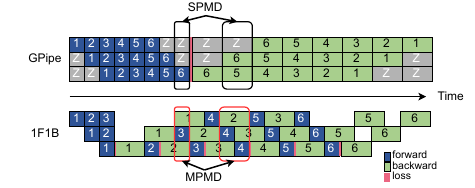}
  \captionof{figure}{
    Comparison between GPipe and 1F1B. In GPipe, at any time, all pipeline-parallel groups
    perform the same computation. Bubbles are implemented as redundant discarded computation~(gray Z blocks).
    In 1F1B, all groups perform different computations.
  }\label{fig:gpipe_1f1b}
  \vspace{-5mm}
\end{figure}

\citeauthor{xu_gspmd_2021} presented a clever encoding of pipeline parallelism as sharding in GSPMD.
Assuming that all the stages have the same dataflow graph and input and output shapes
(i.e. stages are \emph{homogeneous}),
it is possible to ``\emph{stack}'' the weights of the layer and perform all stages in parallel
by sharding the weights on the new leading dimension.
Then, the same computation is applied to a sharded ``state'' buffer for a number of times in a loop,
until all the microbatches have been processed.
During the pipeline bubble iterations, idling actors participate in the computation
(gray Z blocks in \Cref{fig:gpipe_1f1b}) discarding the iteration's result.
After the pipeline loop on the stacked layers, the outputs are used to compute the loss
of the full batch.

Besides being unsuitable for models with non-homogeneous stages,
GSPMD encoding can also negatively impact performance in the following ways:
homogeneous stages forbid using different rematerialization strategies across stages
and strict synchronization at each loop iteration forces all processes to
wait for stragglers.

JAX's automatic differentiation (autodiff) generates a corresponding loop for the backward
pass consuming the activations in reverse order.
After SPMD partitioning, the generated code corresponds to the GPipe schedule.
There is no way for the user to \emph{control} the scheduling of the sections of the
gradient accumulation loop, forgoing potential performance benefits described in~\Cref{subsec:ppsched}.
Although we do not preclude the existence of some program transformations that could encode 1F1B
in the SPMD paradigm under further assumptions, such transformations would be inadequate.
They would fail to respect the true essence and flexibility of pipeline schedules,
which clearly necessitate a MPMD paradigm, where at any time different
actors perform different stages of the loop~(\Cref{fig:gpipe_1f1b} bottom).

\begin{figure*}[ht]
  \centering
  \includegraphics[width=\linewidth]{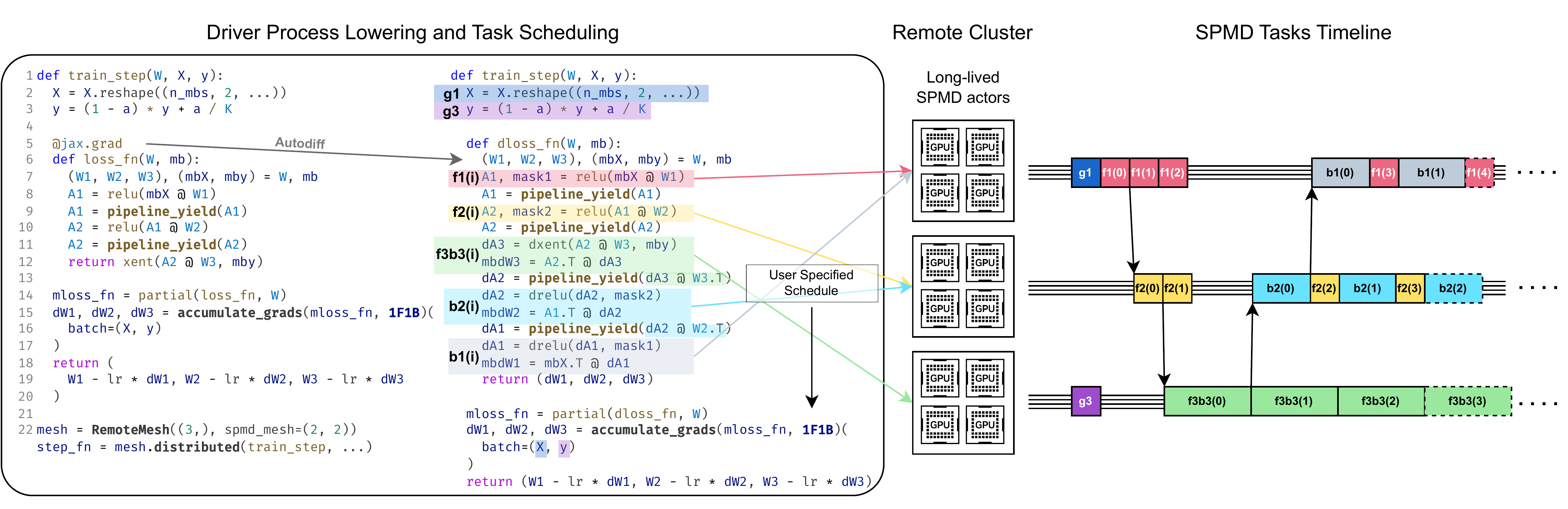}
  \caption{
    System Overview. The left box shows the code in the driver process describing
    the computation and annotating pipeline stage boundaries.
    Auto-differentiation produces additional stages corresponding to the
    ``backward'' computations for the gradients.
    The user specifies a mapping of stages to SPMD actors and a schedule for
    the loop.
    Each call to the \lstinline|step_fn| function schedules tasks
  }\label{fig:main}
  \vspace{-3mm}
\end{figure*}

\section{\sysname Overview}\label{sec:design}

We now describe \sysname, a compiler and runtime for running distributed MPMD computations.
\sysname extends JAX's user-driven SPMD parallelism by introducing task-based temporal parallelism.
To achieve this, it addresses two key challenges:
(1) User-scheduled gradient accumulation:
we introduce a familiar loop construct that integrates user-defined schedules seamlessly in existing code;
(2) Asynchronous task execution: we develop a runtime capable of efficiently executing distributed
task graphs in parallel.

\Cref{fig:main} gives an overview of the compilation and runtime components of the system.
The user's specification of \lstinline|train_step| differs only slightly from a
standard training step in JAX without pipeline parallelism.
Additionally, the code is updated to perform gradient accumulation over the microbatches
with \lstinline|accumulate_grads| (L14) and the model is annotated with auto-differentiable
\lstinline|pipeline_yield| calls marking the end of the current stage (L9, L11).
The \lstinline|distributed| function traces the auto-differentiated training step
into an intermediate representation called Jaxpr,\footnote{\url{https://jax.readthedocs.io/en/latest/jaxpr.html}}
which is transformed and split into multiple tasks by the driver process.
These tasks are then lowered and sent to the respective SPMD actors as specified by the schedule
to be compiled and run on the \emph{remote} devices.
The remote devices are allocated by the driver process by instantiating a \lstinline|RemoteMesh|.
In the example displayed, 3 SPMD actors are provisioned, each with 4 devices configured in a SPMD mesh
shape of \lstinline|(2, 2)|.
\sysname attempts to group devices so that those assigned to an SPMD actor are connected
through a high-bandwidth interconnect.
Each task is lowered and compiled by XLA, leading to the same exact SPMD parallelization
strategy the user would expect as described in~\Cref{subsec:named_axes} within each task.
\sysname infers tasks for all communication and resource management needed
to perform the program's execution, such as send and receive operations and deallocation
of intermediate buffers.
All these tasks are fused into a single MPMD ``program'', so that at each call of the returned
\lstinline|step_fn| function, all the tasks can be dispatched into a single RPC call
per SPMD actor.

In this section we describe key features of \sysname that enable
MPMD pipeline parallelism.

\subsection{Gradient Accumulation Loop}
\Cref{fig:train_loop_jaxpp} highlights in {\color{teal} teal} the changes required to
adopt \sysname in an existing JAX training program.
The model definition can leverage JAX's sharding annotations as shown in~\Cref{subsec:named_axes}.
The code is updated to implement the gradient accumulation loop over the microbatches
with \lstinline|accumulate_grads| (\Cref{line:accumulate}).
The argument to \lstinline|accumulate_grads| is a function (\lstinline|microbatch_grads|)
that given one microbatch produces the gradients and additional metrics
of that microbatch.
Semantically \lstinline|accumulate_grads| will call \lstinline|microbatch_grads|
on each microbatch in \lstinline|batch| and sum the gradients and collect the loss
from each iteration, equivalently to the code below.

\begin{lstlisting}
grads = zeros_like(state.params)
loss = []
for i in range(batch.shape):
  mugrads, muloss = microbatch_grads(batch[i])
  grads += mugrads
  loss.append(muloss)
\end{lstlisting}

The API is configured by default to implement the addition and concatenation operator
on each iteration's output with the loop state.
Internally, the API lowers to a proper structured ``for loop'' with an explicit state
that is updated in the loop body.
This API restriction is intentional to ensure that the provided loop body does
not create dependencies between earlier stages of the current iteration of the loop
with later stages of the previous iteration.

During compilation the gradient accumulation loop is ``unrolled'' into a task
graph that is then scheduled and run on the remote devices.

\begin{minipage}{\linewidth}
\begin{lstlisting}
$\text{@}$shard( () , ("emb", "mlp"), ("mlp", "emb"), ())
def ffn($X$, $W^{(1)}$, $W^{(2)}$):
  $H^{(1)}$ = self.act($X W^{(1)}$)
  $H^{(1)}$ = shard($H^{(1)}$, ("batch", "mlp"))
  $\color{teal} A^{(1)}$ = @jaxpp.pipeline_yield@($\color{teal} H^{(1)}$) $\label{line:pipeline_yield}$
  $H^{(2)}$ = $A^{(1)} W^{(2)}$
  return shard($H^{(2)}$, ("batch", "emb"))

def loss_fn(...):
  # calls ffn
  ...

model, lr_scheduler, state = ...

def train_step(state, batch):
  def microbatch_grads(mubatch):
    # mubatch.shape=(mbsz, *rest)
    muloss, mugrads = jax.value_and_grad(loss_fn)(
      state.params, mubatch
    )
    # muloss.shape=()
    return mugrads, muloss
    #      $\color{violet}+$       $\color{violet}\lVert$

  @schedule = _1F1B(stages=2)@

  # batch.shape=(n_mbs, mbsz, *rest)
  grads, loss = (
    @jaxpp.accumulate_grads(microbatch_grads, schedule) $\label{line:accumulate}$
      (batch)@
  )
  # loss.shape = (n_mbs,)
  new_state = state.apply_gradient(
    grads, learning_rate=lr_scheduler(state.step)
  )

  return new_state, loss

@mesh = RemoteMesh((2,), spmd_mesh=(2, 2))@
jit_train_step = @mesh.distributed@(
  train_step,
  in_shardings=(state_sharding, batch_sharding),
  out_shardings=(state_sharding, None)
)

for batch in dataset:
  state, loss = jit_train_step(state, batch)
\end{lstlisting}
\vspace{-3mm}
\captionof{figure}{Training loop in JaxPP}\label{fig:train_loop_jaxpp}
\end{minipage}

\subsection{Stage Marking}
The user specifies the start and end of ``logical stages'' through
\lstinline|pipeline_yield|.
Any computation arising before the first call
to \lstinline|pipeline_yield| is implicitly scheduled on the first stage,
with each call ``opening'' a new stage.
In \Cref{fig:train_loop_jaxpp},
\sysname ensures that any computation that $A^{(1)}$ (\Cref{line:pipeline_yield}) depends on is scheduled
on the current stage and any computation that depends on $A^{(1)}$ is scheduled
on the next stage.
Note that a stage is simply a unit of computation and is not immediately
bound to a concrete set of execution devices.
\lstinline|pipeline_yield| may be used multiple times to create multiple stages.

Differently from an API where a stage is implemented as a separate function,
marking stages with \lstinline|pipeline_yield| is advantageous in the following ways:
(1) the code change is less disruptive for existing code bases than restructuring
tasks in separate functions and, more importantly,
(2)~signals to the user that stages are entirely dictated by data dependencies.
Indeed, a definition such as \lstinline|x = ...| preceding a \lstinline|pipeline_yield(y)|
is scheduled in the stage preceding the yield only if \lstinline|y| depends on
\lstinline|x|.
Otherwise, \lstinline|x|'s definition operation is scheduled closer to its use,
to minimize communication.
If tasks were defined as functions, autodiff could produce
``gradient merging'' operations (additions) that do not belong to any function
which would have to be added back to one of the tasks.

\subsection{Placement Inference}
Given the \lstinline|pipeline_yield| annotations and the accumulation loop,
\sysname automatically infers data placement for inputs and outputs of the
\lstinline|train_step| function.

In order for pipeline parallelism to work efficiently we have to
ensure that each computation is scheduled at the right pipeline execution
unit where the data are ``pinned'' while at the same time minimizing
communication across actors.
We assume that the loop schedule maps backward computations to the same
actor of the corresponding forward computation.
For example if weights $W_1$ and $W_2$ are placed on a specific actor
then all the computation corresponding to the backward computation for the gradients
of $W_1$ and $W_2$ must be scheduled on the same actor.

We use the following propagation heuristic to schedule operations on a task.
First, a task is formed for each \lstinline|pipeline_yield| operation, comprising
of all computations it depends on.
Then the remaining computations that are not dependencies of any \lstinline|pipeline_yield|
operation are placed on the same task of their operands or a new task.
In the loop body we do not allow any computation replication, and instead
each operation can be assigned to only one task.
This step also infers the placement of inputs and outputs of the \lstinline|accumulate_grads| loop.

Then input placement is propagated to the computation
preceding the pipeline loop, potentially replicating computation
and ensuring that the inferred placement does not overwrite the current placement
and similarly loop output placement are propagated to the computation after
the loop.

\subsection{Weight Sharing and Gradient Accumulation}
In the presence of weight sharing where the same weight is used across multiple
stages as with tied embeddings in Transformer models, multiple
partial gradients are computed from each use which are then
added to form the full gradient, e.g., $g = ((g_1 + g_2) + g_3) + \dots$.
A naive scheduling of such operations would lead to sends and receives
of multiple partial gradients which, for embedding tables, can
easily consist of several Gigabytes of data.

\sysname implements a loop commuting pass which substitutes the carried state of
the cumulative total gradient $g$ with a carried state of the partial
gradients $g_1, g_2, g_3, \dots$ and a final addition operation.
This corresponds to the following rewrite rule. %
\begin{minipage}{\linewidth}\footnotesize
\begin{align*}
  g &= \sum_{i = 1}^{\text{\#microbatches}} \left( g_1^{(i)} + g_2^{(i)} + \dots \right) \\
  &\rightsquigarrow \left(\sum_{i = 1}^{\text{\#microbatches}} g_1^{(i)} \right) + \left( \sum_{i = 1}^{\text{\#microbatches}} g_2^{(i)} \right) + \dots
\end{align*}
\end{minipage}

\section{Scheduling and Runtime}\label{sec:sched}
In the previous section we introduced key Jaxpr transformations and user functions
in \sysname. We now describe the runtime architecture and how tasks are dispatched.

\subsection{Architecture}\label{subsec:scmpmd}
The user program containing the code for training is run in a single
Python process, we call driver or controller, on a host possibly co-located in the datacenter
where training nodes reside.
The controller is responsible for the tracing and transformations described in the previous
section.
Additionally the controller allocates stateful actors managing one or multiple devices
possibly spanning multiple hosts.
We use Ray~\cite{222605} for Remote Procedure Calls (RPCs) and orchestrating worker processes
running XLA computations.
We implement a custom on-device object store on each actor for storing sharded device buffers.
Communication is handled using NCCL P2P operations.

\subsubsection{Single-Controller MPMD}
Using a single-controller model is not strictly necessary for an MPMD implementation of pipeline parallelism,
but it offers several key benefits, such as easier scaling and reduced code complexity.
In contrast, a multi-controller approach can be implemented by replicating the controller logic across all processes,
allowing each process to generate its own local tasks.
However, this method requires users to carefully manage code
sections that need to run in a single-threaded context and manually dispatch processes to ensure correct
placement---for example, keeping model-parallel groups on the same host while distributing pipeline-parallel
groups across different hosts.
With a single-controller model, users can scale from a single-device setup to multiple devices across
hosts with minimal code changes.
The primary effort involves annotating the training step function, simplifying the transition without extensive rewrites.

\subsection{Task Scheduling}

A user can specify a loop schedule by providing a list of tasks for each actor as follows
describing the iteration of the loop run, its type (forward or backward) and the stage index.
\begin{lstlisting}
[
  [ # actor_1
    Task(i=0, ty='fwd', stage=0),
    Task(i=1, ty='fwd', stage=0),
    Task(i=0, ty='fwd', stage=2), ...
  ],
  [ # actor_2
    Task(i=0, ty='fwd', stage=1),
    Task(i=1, ty='fwd', stage=1),
    Task(i=0, ty='fwd', stage=3), ...
  ],
]
\end{lstlisting}

\sysname builds a task graph based on task placement and task dependencies to then
infer allocation, send, and receive operations.
Care has to be taken when generating the local task schedule for each actor,
especially in the generation of send and receive operations.
This is because communication primitives, although asynchronous, still
require send and receive operations to have matching receive and send
operations in the same order respectively among the communicating processes
to prevent potential deadlocks.
Therefore, simply iterating over each local task of an actor and performing receive operations
for the non-local task operands, executing the task and sending the results
immediately as shown in \Cref{fig:deadlock} can potentially result in deadlock.

\begin{minipage}{\linewidth}
\begin{center}
\includegraphics[width=.8\linewidth]{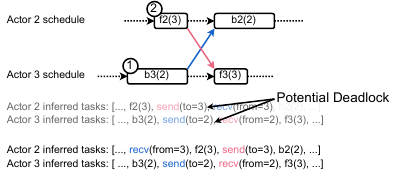}
\end{center}
\vspace{-15pt}
\captionof{figure}{
  Inference of send and receive operations based on uses and definitions
  in the task graph.
}\label{fig:deadlock}
\end{minipage}

Instead, \sysname iterates over the tasks in their topological order and
schedules asynchronous send and receive pairs immediately after the
corresponding task has produced the data to be communicated.
In the example above, after scheduling \lstinline|b3| on actor numbered 3, \sysname immediately
schedules a send and the corresponding receive on the receiving actor.
Since receive operations are asynchronous, the computation of task \lstinline|f2(3)|
is overlapped with the potential prefetching of the data from actor 3 which is
used only later in task \lstinline|b2(2)|.

\subsection{Buffer Deletion}
After generating the local task schedule for each actor, a buffer liveness
pass inserts deletion operations for intermediate buffer.
A buffer that is sent to an actor is tentatively deleted if the corresponding send
operation has completed, otherwise it is tracked into a ``pending deletions'' queue
for later reclamation.
Each scheduled deletion operation checks this queue and deletes previously pending deletions.

\subsection{Task fusion}\label{subsec:schedinstr}
A direct implementation of task dispatch on the driver would perform a separate remote procedure call.
Multiple round trips of ``control'' would lead to poor utilization, especially when running in a loop.
The \lstinline|distributed| annotation fuses all task dispatches into a single RPC call per actor.
All the coordination between multiple actor is resolved by send and receive dependencies only.

\section{Evaluation}\label{sec:evaluation}

In this section, we analyze important performance characteristics of pipeline parallelism as implemented
in \sysname~(\Cref{subsec:characteristics}) and evaluate the performance gains achieved by \sysname
in comparison to other systems that support large language model training with various parallelism
strategies~(\Cref{subsec:perf_comp}).
We conducted our experiments on NVIDIA EOS cluster~\cite{eos_cluster} which is equipped with NVIDIA DGX H100 with
the InfiniBand NDR400 interconnect.
Each node consists of 8 H100 GPUs, each with 80 GB of memory.
We evaluate \sysname on the training of GPT-3 175B~\cite{brown_language_2020-1} and Llama2 70B~\cite{touvron2023llama2openfoundation} at BF16 precision.

\subsection{Performance Characteristics}\label{subsec:characteristics}
In this section, we discuss the performance characteristics of \sysname for specific
configurations, relating them to the design of the system and its potential overheads.

\begin{figure}
  \centering
  \includegraphics[width=.9\linewidth]{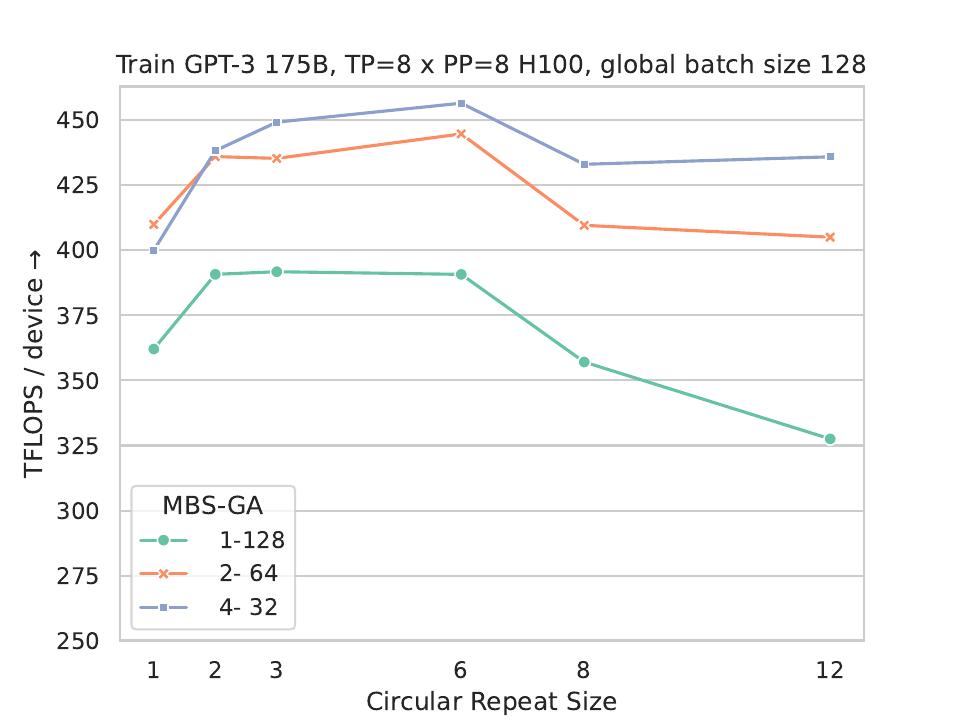}
  \vspace{-4mm}
  \caption{
    Performance of GPT-3 175B training on 64 GPUs with global batch size of 128 instances across
    various configurations for interleaving/circular repeat and microbatch size.
  }\label{fig:mbs_vp}
  \vspace{-4mm}
\end{figure}

\subsubsection{Interleaving and Dispatch Overhead}\label{subsubsec:interleaving}
An important distinction of \sysname over other plain JAX implementations is that
\sysname splits the training step computation into multiple XLA SPMD tasks,
e.g., forward and backward computations for each stage.
This is necessary to implement the various pipeline schedules.
However, it can incur dispatch overheads.
Such overheads can especially be exacerbated when using configurations that try to reduce
pipeline bubbles, such as:
(1) slicing the dataflow graph
into smaller stages and using interleaved schedules~\cite{narayanan_efficient_2021}
(2) slicing the batch into smaller microbatch sizes resulting in more microbatches.

Smaller stages as in (1) increase the number of XLA asynchronous dispatches which have non-negligible cost if the device
work dispatched is too small.
Smaller microbatches as in (2) can lead to poorer kernel-level device utilization and increase the number of
collectives, e.g., the kernel time $t_2$ for one microbatch of size 2 can be smaller than kernel time for 2 microbatches
of size 1 each taking $t_1$ ($t_2 < 2 t_1$).

\Cref{fig:mbs_vp} explores this tradeoff. As shown in the picture increasing the number of circular repeats
of stages, leading to smaller tasks, improves for all cases up to the point when the tasks become
too small and XLA dispatch overheads emerge and P2P latencies start becoming non-negligible.
Increasing the microbatch size increases the bubble time but at the same time reduces
the number of collectives per loop iteration, overall improving performance.

\begin{figure}
  \centering
  \includegraphics[width=.9\linewidth]{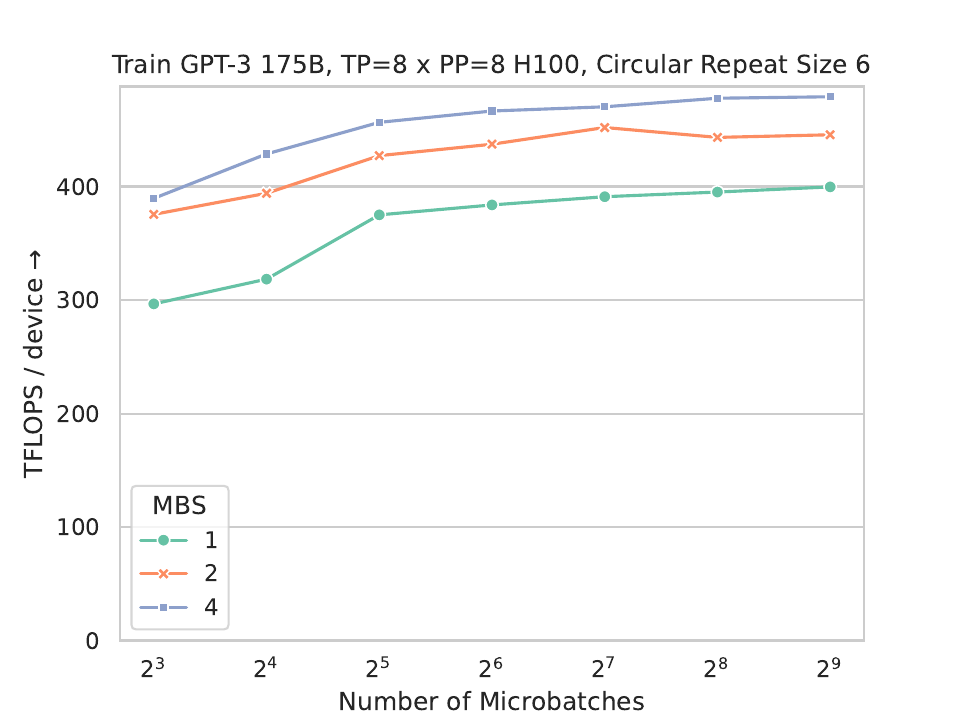}
  \vspace{-4mm}
  \caption{
    Performance of GPT-3 175B training on 64 GPUs with circular repeat size of 6 and
    various combinations of gradient accumulations and microbatch sizes.
  }\label{fig:mbs_ga}
  \vspace{-1mm}
\end{figure}

\subsubsection{Utilization Tradeoff}\label{subsubsec:ga}
Given a model size and a pipeline parallel configuration, it is possible to increase the
overall global batch size either by accumulating over more microbatches at each training step
or scaling in purely data-parallel fashion.
Increasing the number of microbatches is beneficial to minimize the pipeline bubble.
However, given a fixed target number of tokens to train on, it increases end-to-end training time
since more work is done iteratively instead of being parallelized.
At the same time, scaling in data-parallel fashion at low utilization is not cost effective.
\Cref{fig:mbs_ga} shows the utilization achieved by \sysname at different numbers of
microbatches for multiple microbatch sizes.

\subsubsection{Scalability}\label{subsubsec:scaling}
In order to test the scalability of \sysname, we conducted weak scaling experiments on the GPT-3 175B model
by increasing the global batch sizes linearly from 128 to 2048, with 32 microbatches, doubling
the number of GPUs.
8-way tensor parallelism was enabled within each node containing 8 GPUs, and 8-way pipeline parallelism
was enabled across each group of 8 nodes, using Interleaved 1F1B as the schedule and a circular repeat
of size 6.
We instantiate a \sysname actor per node.

\begin{figure}
  \centering
  \includegraphics[width=.9\linewidth]{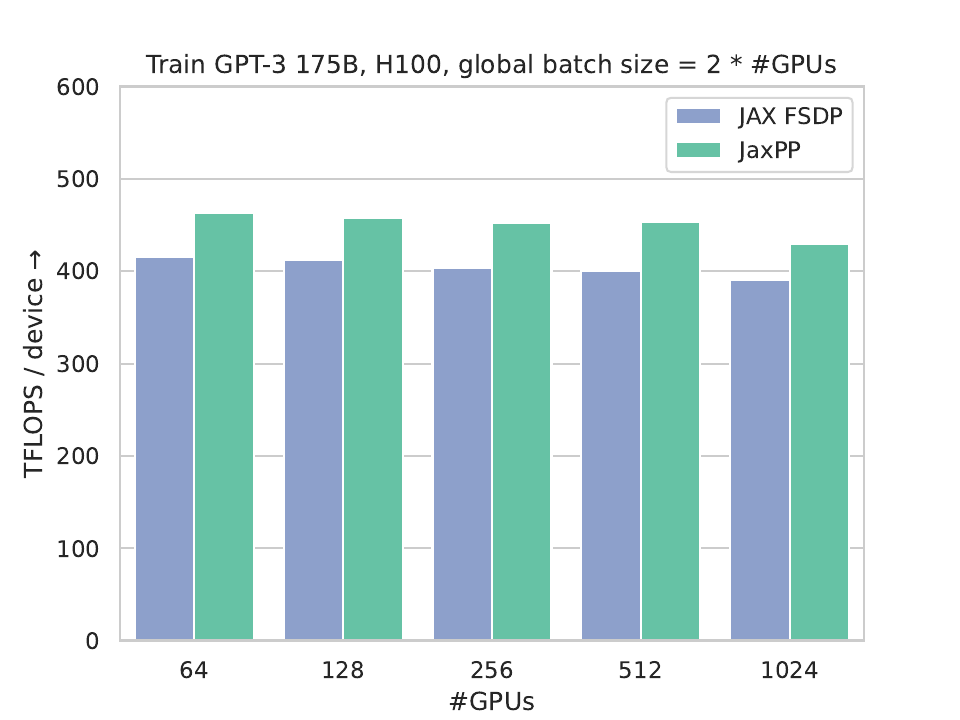}
  \vspace{-5mm}
  \caption{
    \sysname's weak scaling in comparison to a highly optimized JAX FSDP implementation.
  } \label{fig:scaling_gpt3}
  \vspace{-5mm}
\end{figure}

As illustrated in \Cref{fig:scaling_gpt3}, \sysname effectively scales GPT-3 175B training from 64 to 1024
GPUs, achieving a 92.87\% weak scaling efficiency.
This performance is comparable to the 93.97\% efficiency demonstrated by a highly optimized system utilizing
Fully-Sharded Data Parallelism (FSDP) with JAX.
\sysname not only matches the scaling efficiency of JAX FSDP but also delivers higher throughput
and lower end-to-end latency.

\subsection{Training Performance}\label{subsec:perf_comp}

In this section, we compare the performance of model training in \sysname against
the SPMD-based pipeline parallelism solution in JAX and the state-of-the-art implementation found in NeMo~\cite{Harper_NeMo_a_toolkit}.
This comparison aims to validate our claim that \sysname overcomes the limitations of SPMD-based pipeline
parallelism without compromising performance.
Additionally, we discuss how \sysname achieves performance gains in certain scenarios and explore potential
avenues for further improving its performance.

\begin{figure}[ht]
  \centering
  \includegraphics[width=.9\linewidth]{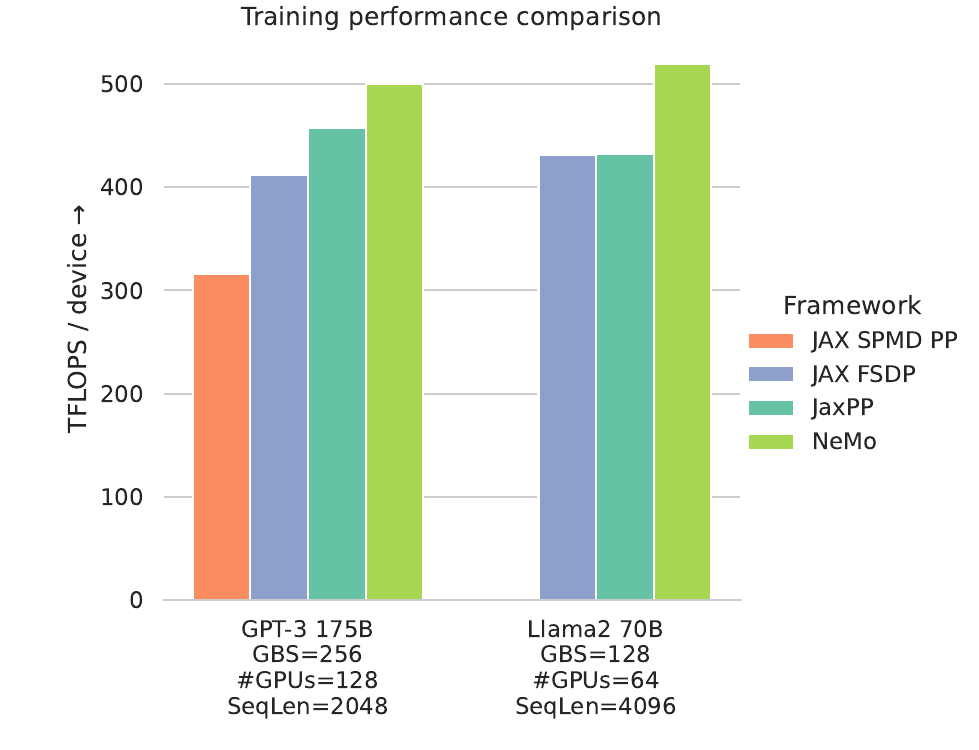}
  \vspace{-5mm} 
  \caption{
    Performance comparison between SPMD pipeline parallelism, \sysname, and NeMo on GPT-3 175B and Llama2 70B.
  } \label{fig:perf_comp}
\end{figure}

As depicted in \Cref{fig:perf_comp}, when training GPT-3 175B on 16 DGX H100 nodes (128 GPUs), \sysname is
44.6\% faster than SPMD pipeline parallelism, achieving 457 TFLOPS/device, while being more expressive
and requiring 1K fewer lines of user code.
Moreover, \sysname improves throughput by $1.11\times$ over JAX's FSDP.
\sysname achieves 91.4\% throughput of NeMo's pipeline parallelism while being entirely model-agnostic.
When training Llama2 70B on 8 DGX H100 nodes (8 GPUs), \sysname demonstrates similar performance as JAX FSDP,
only showing 83.2\% of NeMo's throughput.
We note that NeMo leverages several high-performance kernels that greatly improve end-to-end
performance.
\sysname uses no custom kernels except for the attention APIs from cuDNN~\cite{chetlur2014cudnnefficientprimitivesdeep}.

\begin{table*}[htbp]\footnotesize
  \centering
  \begin{tabular}{cccccccccc}
  \toprule
  \textbf{System} & \textbf{GBS} & \textbf{GA} & \textbf{GPUs} & \textbf{PP} & \textbf{TP} & \textbf{DP} & \textbf{FSDP} &
  \textbf{Step Time (s)} & \textbf{TFLOPS / device}\\
  \midrule
  \rowcolor{gray!10}\multicolumn{10}{c}{\textbf{GPT-3 175B --- BF16 --- Sequence Length 2048}}\\
  \multirow{5}{*}{\textbf{\sysname}}
  &  128 &  32 &   64 &  8 & 8 &  1 &   1 &  9.53 & 462\\
  &  256 &  32 &  128 &  8 & 8 &  2 &   1 &  9.64 & 457\\
  &  512 &  32 &  256 &  8 & 8 &  4 &   1 &  9.74 & 452\\
  & 1024 &  32 &  512 &  8 & 8 &  8 &   1 &  9.71 & 454\\
  & 2048 &  32 & 1024 &  8 & 8 & 16 &   1 & 10.26 & 430\\
  \hline
  \multirow{5}{*}{JAX FSDP}
  &  128 &   1 &   64 &  1 & 1 &  1 &  64 & 10.63 & 415\\
  &  256 &   1 &  128 &  1 & 1 &  1 & 128 & 10.70 & 412\\
  &  512 &   1 &  256 &  1 & 1 &  2 & 128 & 10.91 & 404\\
  & 1024 &   1 &  512 &  1 & 1 &  4 & 128 & 11.01 & 400\\
  & 2048 &   1 & 1024 &  1 & 1 &  8 & 128 & 11.30 & 390\\
  \hline 
  JAX SPMD PP
  &  256 & 128 &  128 & 16 & 4 &  2 &   1 & 13.96 & 316\\
  \hline
  NeMo
  &  256 &  64 &  128 &  8 & 4 &  4 &   1 &  9.78 & 500\\
  \midrule
  \rowcolor{gray!10}\multicolumn{10}{c}{\textbf{Llama2 70B --- BF16 --- Sequence Length 4096}}\\
  \textbf{\sysname}
  &  128 &  16 &   64 &  4 & 8 &  2 &   1 &  8.42 & 432\\
  \hline
  JAX FSDP
  &  128 &   1 &   64 &  1 & 1 &  1 &  64 &  8.44 & 431\\
  \hline
  NeMo
  &  128 &  32 &   64 &  4 & 4 &  4 &   1 &  7.02 & 519\\
  \bottomrule

  \end{tabular}
  \caption{
    Training performance of JaxPP, JAX FSDP, JAX SPMD PP, and NeMo with GPT-3 175B and Llama2 70B.
    Different systems may use different combinations of various parallelism strategies based on their resource 
    requirements and performance characteristics.
  }
  \label{tbl:perf_framework}
  \vspace{-3mm}
\end{table*}

\subsection{Performance Breakdown}\label{subsec:spmd_vs_jaxpp}

\begin{figure}
  \centering
  \includegraphics[width=.9\linewidth]{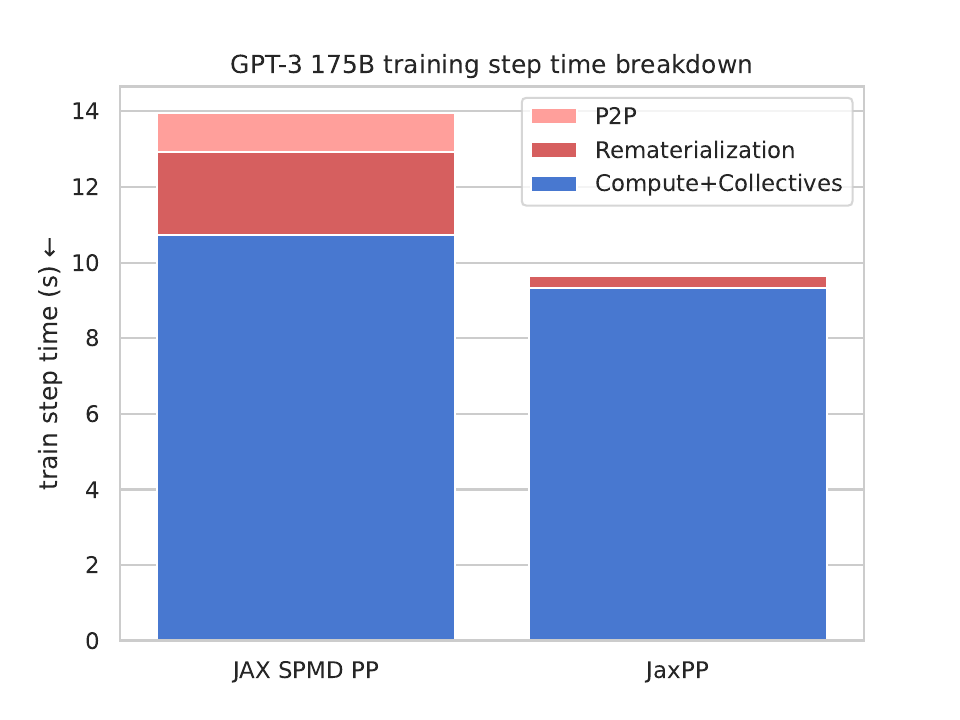}
  \vspace{-5mm}
  \caption{
    Overhead of JAX SPMD PP compared to \sysname.
    Rematerialization cost and asynchronous point-to-point send and receive operations account for the majority of
    the performance differences.
  }\label{fig:breakdown}
  \vspace{-5mm}
\end{figure}

To understand the sources of performance gains achieved by \sysname over SPMD pipeline parallelism on GPT-3 175B,
we present \Cref{fig:breakdown}.
The most significant factor is the rematerialization cost.
SPMD pipeline parallelism employs the GPipe schedule, which has high memory demands, whereas JaxPP utilizes
the Interleaved 1F1B schedule, which requires less memory.
This difference impacts the need for rematerialization, subsequently affecting the overall training step time by $\approx20\%$.
Additionally, \sysname further reduces overhead by overlapping point-to-point send and
receive operations, in contrast to their synchronous counterpart.

\section{Related Work}\label{sec:related}

There are numerous works to facilitate scaling the training of large models~\cite{shoeybi_megatron-lm_2020,10.1145/3394486.3406703,liang2024torchtitanonestoppytorchnative,jiang_megascale_2024}.
Here we discuss systems that are closest to \sysname and explain key design differences.

Alpa~\cite{zheng_alpa_2022} is a system for parallelizing large deep learning models, supporting pipeline parallelism.
Similarly to \sysname, Alpa implements an MPMD runtime on top of JAX/XLA and orchestrates the execution of SPMD tasks.
Nonetheless, Alpa's main focus is automatically inferring the best optimal parallel strategy with respect to a cost
model.
\sysname differs from Alpa in the following ways: it focuses on providing a flexible interface to let the users
control parallelism instead of automating parallelism,
no different from sharding annotations, greatly reducing compilation time;
it does not fork JAX or XLA;
it supports user-extensible stage execution mapping such as Interleaved 1F1B~\cite{narayanan_efficient_2021}.

Pathways~\cite{barham_pathways_2022} is a single-controller distributed dataflow runtime for
machine learning workloads. While some implementation details such as parallel
dispatch and MPMD support are shared between Pathways and \sysname, Pathways is fine-tuned
to time sharing and multiplexing tasks, while \sysname focuses on long-running training jobs,
where resources such as memory, GPU, and interconnect bandwidth are fully allocated to the training job.

Finally, many recent works have proposed novel pipeline schedules for specific scenarios and
new applications~\cite{lamy-poirier_breadth-first_2022,295595,298703,qi2024zero}.
Although we focused on practical applications of traditional schedules,
\sysname has all the features needed to support these novel schedules.

\section{Conclusion and Future Work}\label{sec:conclusion}
We presented \sysname, a system for implementing and efficiently running distributed dataflow
computations.
By extending the SPMD programming model of JAX with temporal parallelism,
we showed that \sysname provides a flexible environment for easily scaling training of deep learning
models with pipeline parallelism.
While the implementation presented here builds on top of JAX and XLA,
the same core ideas can be leveraged to implement similar transformations as an
MLIR~\cite{9370308} dialect and build a MPMD runtime on other technologies.

\section*{Acknowledgements}
We extend our gratitude to Jonathan Dekhtiar, Nitin Nitin, Abhinav Goel,
and Haixin Liu for their invaluable and continuous feedback.
Our thanks also go to Sangkug Lym, Tejash Shah, and Santosh Bhavani for their assistance
in the evaluation of \sysname.
Finally, we are grateful to Shriram Janardhan, Jeremy Wilke, Jaroslav Sevcik, Georg Stefan Schmid,
and Deepak Narayanan for their insightful and thought-provoking discussions.

\bibliography{jaxpp}
\bibliographystyle{mlsys2025}

\end{document}